\documentclass[fleqn,usenatbib]{mnras}

\usepackage[T1]{fontenc}

\DeclareRobustCommand{\VAN}[3]{#2}
\let\VANthebibliography\thebibliography
\def\thebibliography{\DeclareRobustCommand{\VAN}[3]{##3}\VANthebibliography}

\usepackage{graphicx}	
\usepackage{amsmath}	
\usepackage{amssymb}	
\usepackage{xspace}
\usepackage{enumitem}
\newcommand{\SFE}{\ensuremath{\epsilon_{SF}}\xspace}
\newcommand{\LMSR}{\ensuremath{\Lambda_\text{MSR}}\xspace}
\newcommand{\afg}{\ensuremath{\alpha_\text{1GEN}}\xspace}
\newcommand{\asg}{\ensuremath{\alpha_\text{2GEN}}\xspace}

\usepackage{newtxtext,newtxmath}

\title[30 Doradus by $N$-body \& \textsc{warpfield} ]{30 Doradus, the double stellar birth scenario by $N$-body \& \textsc{warpfield} clouds}

\author[R. Dom\'inguez et al.]{R. Dom\'inguez$^1$\thanks{E-mail:raul.dominguez@stud.uni-heidelberg.de},
 Eric W. Pellegrini$^1$, Ralf S. Klessen$^{1,2}$, Daniel Rahner$^1$.
\\$^1$ Universit\"at Heidelberg, Zentrum f\"ur Astronomie, Institut f\"ur Theoretische Astrophysik, Albert-Ueberle-Stra{\ss}e 2, 69120 Heidelberg, Germany
\\$^2$ Universit\"at Heidelberg, Interdisziplin\"ares Zentrum fur Wissenschaftliches Rechnen, Im Neuenheimer Feld 205, 69120 Heidelberg, Germany
  }
\begin{document}
\date{Accepted -----. Received -----; in original form -----}

\pagerange{\pageref{firstpage}--\pageref{lastpage}} \pubyear{2022}

\maketitle

\label{firstpage}
  
\begin{abstract}
We study the evolution of embedded star clusters as possible progenitors to reproduce 30 Doradus, specifically the compact star cluster known as R136 and its surrounding stellar family, which is believed to be part of an earlier star formation event.
We employ the high-precision stellar dynamics code  \textsc{Nbody6++GPU} to calculate the dynamics of the stars embedded in different evolving molecular clouds modelled by the 1D cloud/clusters evolution code \textsc{warpfield}. We explore clouds with initial masses of $M_\text{cloud}=3.16 \times 10^{5}$ M$_\odot$ that (re)-collapse allowing for the birth of a second generation of stars. We explore different star formation efficiencies in order to find the best set of parameters that can reproduce the observation measurements. 
Our best-fit models correspond to a first stellar generation with masses between $1.26\times~10^4$~-~$2.85\times~10^4$~M$_\odot$ and for the second generation we find a $M \approx 6.32\times 10^4$ M$_\odot$. Our models can match the observed stellar ages, cloud shell radius, and the fact that the second generation of stars is more concentrated than the first one. This is found independently of the  cluster starting initially with mass segregation or not. By comparing our results with recent observational measurements of the mass segregation and density profile of the central zone we find close agreement, and thus provide supporting evidence for a centrally focused (re)-collapse origin to the multiple ages.
\end{abstract}


\section{Introduction}
\label{sec:intro}
Young star clusters are usually found embedded in molecular clouds from which they were recently born. The surrounding gas is expelled by feedback, in the forms of  ultraviolet radiation, massive stellar winds from OB stars, and/or supernovae (SNe) explosion. The star clusters lose gravitational potential which is most important in determining the dissolution into the field \citep[see e.g..][]{1978A&A....70...57T, 1980ApJ...235..986H,1984BAAS...16..409M, 1997MNRAS.284..785G, 2000ApJ...542..964A, 2001MNRAS.323..988G, 2003MNRAS.338..665B, 2003MNRAS.338..673B, 2005ApJ...630..879F, 2006MNRAS.369L...9B, 2007MNRAS.380.1589B,  2011MNRAS.414.3036S, 2016MNRAS.460.2997L, 2017A&A...600A..49B, 2017ApJ...838..116F, 2018MNRAS.476.5341F, 2017A&A...605A.119S, 2018ApJ...863..171S, 2020IAUS..351..507S}

In this scenario, the amount of feedback is usually assumed to be strong enough to completely disrupt the molecular cloud and as a consequence prevent any further star formation \citep{2011ApJ...729..133M,2010ApJ...709...27W}. Another possible scenario is positive feedback. As the energy and momentum inserted is pushing out the cloud into a shell-like structure the corresponding density might locally trigger collapse and thus another star-formation  event \citep{2012ApJ...744..130K}. 

\citet{2017MNRAS.470.4453R} introduced the code \textsc{warpfield} (\textbf{W}inds \textbf{A}nd \textbf{R}adiation
\textbf{P}ressure: \textbf{F}eedback \textbf{I}nduced \textbf{E}xpansion, col\textbf{L}apse and \textbf{D}issolution), which models another scenario called failed feedback, where the molecular cloud can (re)-collapse. This semianalytic 1D model for isolated massive clouds with masses $\geq10^5 \text{M}_\odot$ describes the dynamics and structure of the expanding or contracting shell due to winds, SNe, radiation pressure, and gravity. This approach allows us to explore a large range of parameters of star formation efficiency (\SFE), density ($n_0$), and metallicity in a reasonable quantity of CPU-time. A new version of the code was introduced in \citet{2019MNRAS.483.2547R} where the treatments of the thermal evolution of the gas were improved.

30 Doradus, located in the Large Magellanic Cloud (LMC), is a massive star forming region. In its center, the cluster NGC~2070 hosts a younger massive subcluster, R136. It appears that older stellar population in NGC~2070 did not produce enough feedback to take apart its parental molecular cloud, which could retain or re-accrete part of its mass, and form R136 as a massive second generation cluster. The last has been supported by simulation, e.g., \citet{2017MNRAS.465.1375S} showed that under dense conditions ($n \gtrsim 10^5 \ \text{cm}^{-3}$ in a cloud of $10^6$ M$_\odot$), the feedback produced by stellar winds may not be as stronger as it is needed to disperse the cloud. There are other massive young clusters, which show evidence for multiple generations of stars, e.g.\ Sandage-96 exhibits a bimodal age separation of at least 10 Myr \citep{2005ApJ...626L..49P,2007ApJ...659L..41P,2009ApJ...695..619V} or the Orion nebula cluster with an even smaller age spread of less than 1 Myr  \citep{2017A&A...604A..22B}.

The best evidence for multiple stellar generations in compact star clusters comes from the observations of globular clusters \citep[see e.g..][]{2009A&A...505..117C}. These could be a result of the (re)-collapse of gas ejecta from older generation asymptotic giant branch stars \citep{2008MNRAS.391..825D}, fast-rotating massive stars \citep{2007A&A...464.1029D} or interactive massive binaries \citep{2009A&A...507L...1D}. However, these scenarios typically predict that the second generation of stars is much less massive than the older generation, specifically for a small age difference, and so they are not applicable to  30 Doradus where both stellar populations have roughly equal mass.

We structured the paper as follow. First, in Section 2, we explain the method and parameter space where we develop our study. In Section 3, we investigate 30 Doradus scenario and match the observables with a different range of \textsc{warpfield} models and $N$-body simulations. In Section 4, we discuss the results and we conclude in Section 5.

\section{Method and Initial Conditions}

\subsection{Properties of 30 Doradus}
 The main cluster in the 30 Doradus region, NGC~2070, contains two stellar generations. The older population has an age of $\sim$ 3-7 Myr \citep{Brandl1996,Walborn1997,Selman1999,Sabbi2012,Cignoni2015} and the younger population $\sim$ 0.5-2 Myr  which also appears to be more concentrated towards the centre \citep{Massey1998,Selman1999,Sabbi2012,Cignoni2015,2020MNRAS.499.1918B,2022arXiv220211080B} called R136 or formally known as RMC 136. The masses of the clusters are poorly constrained. R136 has a range of mass between 2.2x10$^4$-1x10$^5$ M$_\odot$ \citep{Hunter1995,Andersen2009,Cignoni2015} and the whole cluster NGC~2070 6.8x10$^4$-5x10$^5$ M$_\odot$ \citep{Selman1999,Bosch2001,Bosch2009,Cignoni2015}. In this zone is observed ionized gas which forms bubbles containing hot, X-ray emitting gas \citep{2006AJ....131.2164T}. \citet{Pellegrini2011} using [SII]/H$\alpha$ observations showed that the H II region around NGC~2070 has the shape of a hemispherical bowl. The whole sphere has a radius of 40 - 60 pc and R136 has an offset approximately 12 pc from its centre. The shell radius surrounding R136 as the centre is $\sim$ 30 - 70 pc. We summarize these values in Tab. \ref{tab:tabpar}.

\subsection{Modeling approach}
Our goal is to find the cloud-cluster parameter space capable of reproducing the observables of 30 Doradus sensitive to cluster evolution. To address this problem, we study a range of molecular clouds and cluster masses, resulting in different \SFE. The evolution of the clouds is followed using the code \textsc{warpfield} 2.1 \citep{2017MNRAS.470.4453R}. As the clouds expand, the gravitational potential is changing, which is introduced into the $N$-body calculation of the stellar dynamics as a time-evolving external potential. The dynamics of the star clusters is followed using the code \textsc{Nbody6++GPU} \citep{2015MNRAS.450.4070W} modified for our purpose in order to read in information from the \textsc{warpfield} code. We note that \textsc{warpfield} calculates the overall feedback produced by a star cluster located in the centre of the cloud. The energy injected from the stars to the cloud produces its expansion, resulting in one of the following outcomes:
\begin{enumerate}[left= 0pt]
\item The cluster inject enough feedback dispersing the cloud.
\item The cluster does not inject enough feedback and after an initial period of expansion, gravity overtakes and the cloud collapse again and gives birth to a new stellar generation. 
\item The subsequent evolution can follow (i) or (ii), which means the process could be repeated multiple times leading to the formation of multiple stellar populations until the cloud is finally dispersed. 
\end{enumerate}

Using \textsc{warpfield}, we create clouds with masses of $3.16~\times~10^{5}$~ M$_\odot$ following uniform profiles which host star clusters of different masses. From the observational data, the R136 appears to be more massive than the old cluster. We fix the new cluster to have a value of \SFE = 0.20 and we try for older stellar component \SFE between 0.01 and 0.10. To emulate our star clusters, we follow Plummer density profiles with $R_\text{pl}$ = 1 pc and the stellar mass is changed to achieve the \SFE required. We randomly create 10 different Plummer distributions using \textsc{mcluster} \citep{2011MNRAS.417.2300K} following a \citet{2001MNRAS.322..231K} initial mass function (IMF). The masses are randomly located along the different Plummer distributions to obtain two samples. Each sample consists in 10 clusters with mass segregation and 10 non-segregated. To be consistent with \textsc{warpfield} calculations, we are not using the stellar evolution features from \textsc{Nbody6++GPU}. We are evolving massive stars until they reach their maximum ages according to \citet{2012A&A...537A.146E} as \textsc{warpfield} follows. The size of the cluster is a free parameter for \textsc{warpfield}, which determines the radial 1d feedback. We use $R_\text{pl}$ = 1 pc as is commonly assumed for young clusters in the range of mass used in this work \citep{2016A&A...586A..68P}.

\begin{table}
\caption{Summary of observational parameters to match with our models.}
\label{tab:tabpar}
\centering
\begin{tabular}{|c|c|}
\hline
Observable & Value \\ \hline\hline
Age first stellar generation & 3-7 Myr \\
Age second stellar generation & 0.5-2 Myr \\
Mass second stellar generation & 2.2x10$^4$-1x10$^5$ M$_\odot$ \\
Total mass NGC~2070 & 6.8x10$^4$-5x10$^5$ M$_\odot$\\
Shell radius & 30 - 70~pc \\
\hline\hline
\end{tabular}
\end{table}

From \textsc{warpfield} outputs, we obtain the cloud boundary for different times. These boundary conditions set the initial conditions the photoionized/photo dissociation region/cloud interface. Using {\sc CLOUDY} \citep{2017cloudy} these initial conditions result in a radial density profile for every time step. To calculate the potential and forces, we use for each of the snapshots Poisson's equation:
\begin{equation}
\nabla^2 \phi(r) = 4\pi G \rho(r),
\end{equation}
where $\phi(r)$ is the radial gravitational potential produced by the cloud, G the gravitational constant and $\rho(r)$ is the radial density profile obtained from cloudy reduction. From the potential calculation, we obtain the radial force $F(r)$ as:
\begin{equation}
F(r)=-\frac{d}{dr}\phi(r).
\end{equation}
For each case, we use 10 different statistical realisations of the star cluster following a Plummer density distributions, all starting in virial equilibrium including the gas. 
For embedded star clusters the virial ratio $\alpha$ is used, which is defined as:
\begin{equation}
    \alpha=\frac{T}{|\Omega|},
\end{equation}
where $T$ refers to the total kinetic energy, and $\Omega$ is the total gravitational potential. A value of $\alpha=0.5$ means the embedded star cluster is in virial equilibrium, $\alpha<0.5$ means contraction and  $\alpha>0.5$ means expansion. After an exploration of different $\alpha$ states for the second generation of stars, we report results from $\alpha=0.3$ which can reproduce closely the observations of R136 presented by \citet{2021A&A...649L...8K} (hereafter K2021).

\subsection{Analysis}
One key parameter characterizing the dynamical state of a star cluster is the level of mass segregation, which we quantify using the ``mass segregation ratio'' parameter (\LMSR) introduced by \citet{2009MNRAS.395.1449A}. It is defined as:
\begin{equation}
   \LMSR=\frac{\left<l_{\text{norm}}\right>}{l_{\text{massive}}} \pm \frac{\sigma_{\text{norm}}}{l_{\text{massive}}}.
\end{equation}
For this parameter, a value of  \LMSR $ \sim $ 1 indicates no mass segregation, i.e., low and high mass stars are similarly distributed.  \LMSR $ \gg $ 1 indicates strong mass segregation, i.e., massive stars are located close to each other and \LMSR $ < $ 1 means inverse mass segregation, i.e., high mass stars are more dispersed than the rest of the cluster. We compute {\LMSR} for all stars in the system and for the first and second stellar generations separately.

The procedure to match with the observable of NGC~2070 followed in this work is summarized as:
\begin{enumerate}[left= 0pt]
    \item We let evolve an initial $N$-body cluster (1GEN), in equilibrium with the gas (\afg $=0.5$), until the moment when {\textsc{warpfield}} indicates that there is a second starburst ((re)-collapse).
    \item We stop the simulation and we add a second $N$-body cluster (2GEN).
    \item  We scale the velocities of the stars for 2GEN to get \asg~$=0.3$.
    \item We continue the simulation until reach 8 Myr, which is already 1 Myr older than the current age of NGC~2070.
\end{enumerate}
We use 5 different Plummer distributions to represent the 1GEN and another 5 for the 2GEN. In this study we consider the two cases:  both stellar generations either start with mass segregation or without. We compare the central distance for the massive stars and the Lagrangian radii for each generation. We also study \LMSR parameter as a function of simulation time. We are looking for simulations that evolve to produce the observable values in Tab. \ref{tab:tabpar}, and a  \LMSR $ <  1$ when the massive stars of 1GEN and 2GEN are compared, i.e., the older massive stars more dispersed than the younger as NGC~2070 exhibits. For all parameters, we show the average of 5 different realizations.

\begin{table}
\caption{Summary of parameter space explored in this work. The first column shows the initial density of the clouds which initially have a mass of $3.16x10^5$ M$_\odot$. The \SFE for the embedded star clusters are shown in the second column and the time when the clouds (re)-collapse are shown in column third. The temporal duration over which the models match with observations is shown in column fourth. The average separation of the central distance for massive stars between generations at the moment of the match is shown in columns five and sixth for simulations with mass segregation and without, respectively.}
\label{tab:tabsims1}
\centering
\resizebox{\columnwidth}{!}{
\begin{tabular}{|c|c|c|c|c|c|c|}
\hline
$n_0$&\SFE$_1$ & (re)-collapse & $\Delta$ time &\multicolumn{2}{|c|}{$D$[1GEN] - $D$[2GEN] (pc)}\\
 ($\text{cm}^{-3}$) &\SFE$_2$ & time (Myr) & (Myr)& SEG & NOSEG\\
\hline
6000& 0.04-0.20 & 2.63 & 0.60 & 2.93 $\pm$ 0.61& 3.37 $\pm$ 0.58\\
 6000& 0.05-0.20 & 3.35 & 0.70 & 3.28 $\pm$ 0.73 & 3.60 $\pm$ 0.92\\[4 pt]

 7000& 0.05-0.20 & 2.62 & 0.50 & 3.60 $\pm$ 0.39& 3.61 $\pm$ 0.73\\
 7000& 0.06-0.20 & 3.38 & 0.70 & 2.78 $\pm$ 0.44& 3.76 $\pm$ 0.64\\[4 pt]

 8000& 0.05-0.20 & 2.21 & 0.50 & 3.69 $\pm$ 0.46& 3.72 $\pm$ 0.65\\
 8000& 0.06-0.20 & 2.62 & 0.50 & 3.09 $\pm$ 0.52& 3.19 $\pm$ 0.41\\
 8000& 0.07-0.20 & 3.40 & 0.70 & 2.59 $\pm$ 0.46& 2.48 $\pm$ 0.93\\[4 pt]

 9000& 0.04-0.20 & 1.76 & 0.40 &3.79 $\pm$ 0.46& 3.72 $\pm$ 0.62\\
 9000& 0.05-0.20 & 1.95 & 0.40 &3.85 $\pm$ 0.44& 3.91 $\pm$ 0.45\\
 9000& 0.06-0.20 & 2.22 & 0.50 & 3.13 $\pm$ 0.36& 3.10 $\pm$ 0.38\\
 9000& 0.07-0.20 & 2.63 & 0.50 & 2.53 $\pm$ 0.63& 2.73 $\pm$ 0.60\\
 9000& 0.08-0.20 & 3.48 & 0.70 & 1.94 $\pm$ 0.31& 3.12 $\pm$ 0.70\\[4 pt]

 10000& 0.06-0.20 & 1.94 & 0.40 & 3.45 $\pm$ 0.45& 3.36 $\pm$ 0.44\\
 10000& 0.07-0.20 & 2.22 & 0.40 & 2.98 $\pm$ 0.48& 3.53 $\pm$ 0.53\\
 10000& 0.08-0.20 & 2.65 & 0.50 & 2.17 $\pm$ 0.59& 2.95 $\pm$ 0.51\\
 10000& 0.09-0.20 & 3.61 & 0.70 & 1.97 $\pm$ 0.59& 2.86 $\pm$ 0.71\\
\hline
\end{tabular}}
\end{table}

\begin{figure}
\includegraphics[width=\columnwidth]{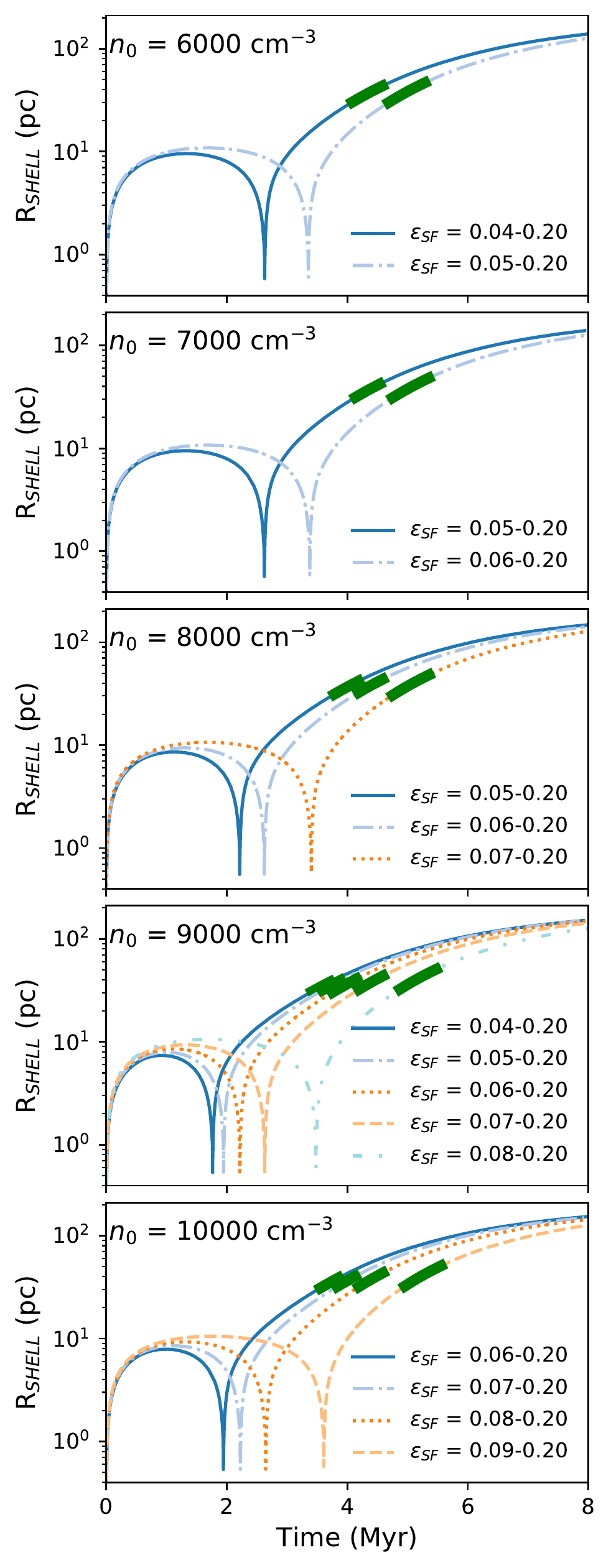}
\caption{Shell radius evolving in time from \textsc{warpfield} models. Different colours and symbols represent the respective \SFE. The ranges of time when the models match with the observable are highlighted by a green thick line.}
\label{fig:rWF55}
\end{figure}   

\section{Results}
\label{sec:results}
\subsection{WARPFIELD clouds}
\label{sec:WFclouds}
We first constrain our parameter space by finding \textsc{warpfield} clouds which can reproduce the ages of the two stellar populations and the shell radius. We explore different values of \SFE from 0.01 until 0.10 for the first cluster and a fixed \SFE = 0.20 for the second star cluster. These choices allow us to match the observed mass of R136. We summarize the successful \textsc{warpfield} models in Tab. \ref{tab:tabsims1} where the first and second columns indicate the initial density and the \SFE pair, respectively. The shell radii evolution for each of the cases in our parameter space obtained from \textsc{warpfield} simulations are shown in Fig. \ref{fig:rWF55}. Every panel shows a cloud with different initial density. We have for every initial density more than one \SFE pair which are represented with different line styles. Initially, the clouds expand due to stellar feedback exerted by the central cluster. After that, depending on the initial density and the cluster mass, the shell radii reach a maximum followed by a (re)-collapse. The (re)-collapse times are shown in Tab. \ref{tab:tabsims1}, column third. The moment of (re)-collapse increases as we use larger \SFE for each cloud. This is expected as a more massive cluster keeps the cloud expansion for a longer time due to higher feedback. After the second starburst, the shells expand again and for all cases, the expansions continue until we stop the simulation. We highlight with a thicker green line the zone where the stellar ages and the shell radius match the observables (see Tab. \ref{tab:tabpar}). For all cases, the left sides of the matching zones start when the minimum shell radius is found ($\sim 30$ pc) and the right limit when the 2GEN maximum age is reached. The temporal duration ($\Delta$) of these zones are summarized in Tab. \ref{tab:tabsims1}, column fourth with values between 0.4 and 0.7 Myr. Two clusters together produce a faster expansion of the shell, as a larger amount of feedback is added to the cloud. If less massive 2GEN clusters (\SFE < 0.20) are taken into account, these zones are much shorter as the shells need more time to reach the minimum size, approaching or even passing the maximum 2GEN age. The inclusion of mass segregation does not change the \textsc{warpfield} cloud evolution, as the 1D model simply assumes all feedback is injected from the cluster centre. This is a appropriate assumption as the shell radius  exceeds the cluster radius during most evolutionary phases, except at the very end of (re)-collapse when a new stellar generation is formed (Domínguez et al. 2022, submitted to MNRAS).

\begin{figure*}
\includegraphics[width=0.99\textwidth]{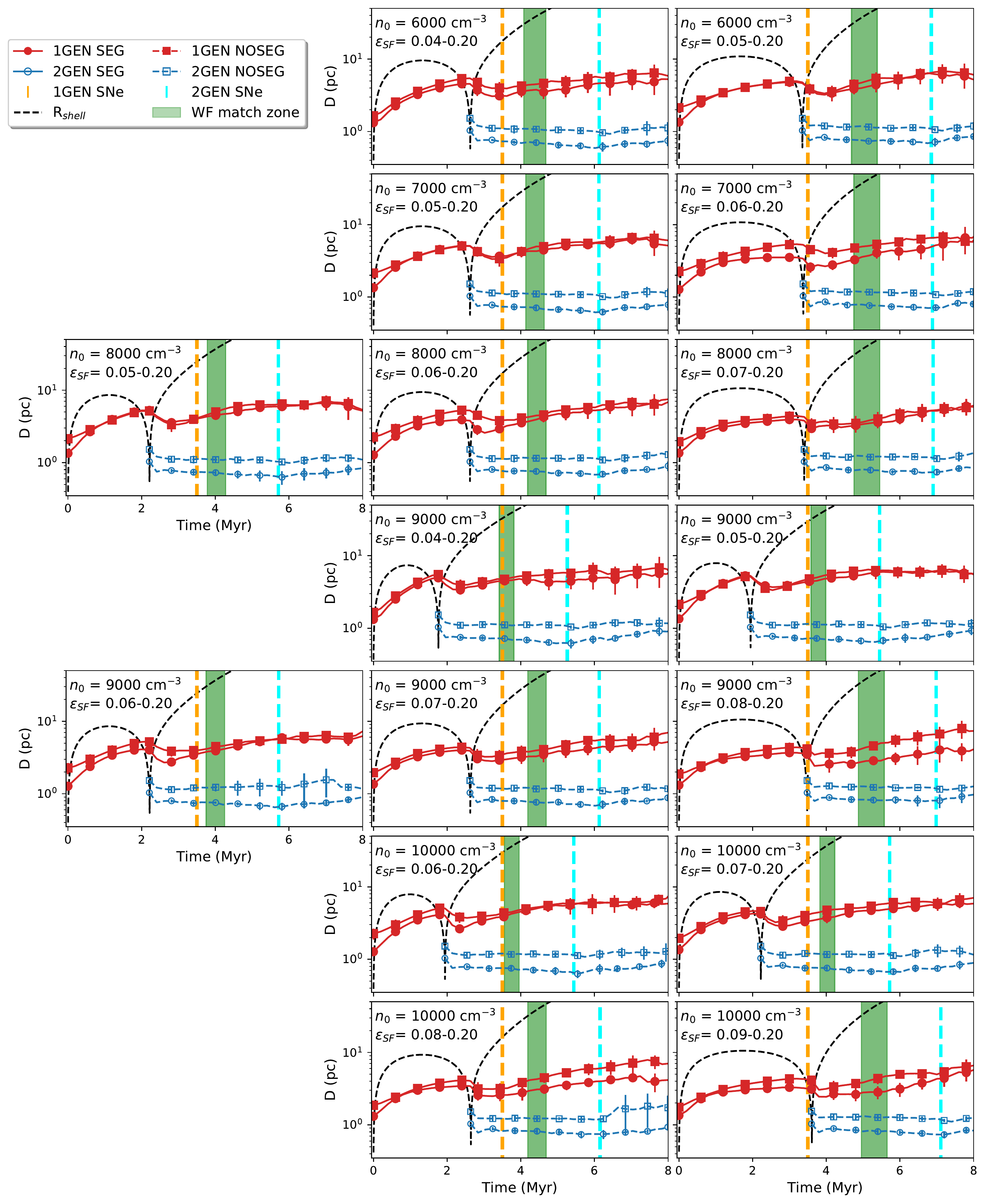}
\caption{Central distance of massive stars ($M > 20$  $\text{M}_\odot$) vs time. The first generation (1GEN) is denoted by red filled symbols and a red solid line. The second generation (2GEN) is denoted by blue empty symbols and dashed blue lines. If the simulations start with mass segregation (SEG) or not (NOSEG) is represented by circles or squares, respectively. The Black dashed line is the shell radius. The green zone indicates where the ages of 1GEN and 2GEN match with the shell radius. The times when SNe start for each generation are denoted by orange and cyan vertical dashed lines, respectively. The information of the initial cloud density ($n_0$) and star formation pairs (\SFE) are given in every panel.}
\label{fig:rmasm55SEGNOSEG}
\end{figure*}

\subsection{Massive stars}
\label{sec:masstars}
In Fig. \ref{fig:rmasm55SEGNOSEG}, we compare the central distance evolving in time for massive stars ($M>20 \text{M}_\odot$) by stellar generation. We show the central distance evolution for each of the cases weighted by its luminosity according to \citet{2011A&A...535A..56G} to achieve more specific information about the location of the most massive stars which predominate in brightness and quantity of feedback. Simulations starting with mass segregation (SEG) are shown with circles and with no mass segregation (NOSEG) are denoted by squares.  The evolution for 1GEN is denoted by red filled symbols and 2GEN by blue empty symbols. The green zone is the matching zone described in Sec. \ref{sec:WFclouds} and the shell radius is represented by a black dashed line. We also indicate when the first SN occurs for each generation with orange and cyan dashed lines, respectively. This is $t \sim t_0 + 3.5$ Myr, where $t_0 = 0$ Myr for 1GEN and for 2EGN is the time when the cloud (re)-collapses. The clusters cover a range of masses for 1GEN of $1.30 \times 10^4 \ \text{M}_\odot \leq M_\text{1GEN} \leq 2.85 \times 10^4 \ \text{M}_\odot$. On the low mass limit, the small number of stars is not enough to cover the whole IMF mass range and this is only complete for \SFE $\geq 0.06$. On the other hand, for 2GEN we have $M_\text{2GEN} \approx 5.90 \times 10^4$, which completes the IMF sample.

For SEG simulations, we observe that 1GEN massive stars (solid line, filled red circles) reach outer positions due to cloud expansion affecting them. At the moment of the (re)-collapse, a strong gravitational potential on the centre is produced due to the high density of the cloud and 1GEN expansion is reversed. After the starburst and with the second cloud expansion, the massive stars are found travelling inward toward the centre. A small contraction of the distribution of the older stars is observed, which is followed by a steady expansion until the end of the simulation. We do not observe a clear effect of the SNe as they mostly start with the clusters already in expansion. On the other hand, 2GEN massive stars (dashed line, empty blue circles) start more concentrated, as described by their initial mass configuration. For 2GEN clusters birthed with  $\alpha=0.3$,  the stellar distributions contract and stabilize in a more concentrated state compared to 1GEN, until the moment when SNe start.  At 8 Myr, we observe the expansion of these younger stars is less than the older, practically ignoring the change in gravitational potential, and remaining always more concentrated than 1GEN.

For NOSEG simulations, 1GEN massive stars (solid line, filled red squares) begin, as expected, less concentrated than SEG clusters. We note, however, that they show similar dynamical evolution to clusters with initial segregation. The same description can be applied for 2GEN  (dashed line, empty blue squares) with central distances always larger than each SEG pair. The effect of the SNe is less visible and after this point, SEG and NOSEG curves approach common values at late times. As before, we observe mass segregation between the different aged populations.

At the moment when the different curves cross the green zone, the older massive stars are more expanded than the younger massive stars. They show different separations and their values are summarized in Tab. \ref{tab:tabsims1} in columns four and five for SEG and NOSEG cases, respectively. The separation cover values between 1.68-4.10 pc. No trend is observed for the different initial density clouds and \SFE pairs. Our best and worst models are for SEG sample correspond to $n_0 = 9000$ cm$^{-3}$ with \SFE = 0.04-0.20 and $n_0 = 10000$ cm$^{-3}$ with \SFE = 0.09-0.20, respectively. For NOSEG sample, our best model is for $n_0 = 6000$ cm$^{-3}$ with \SFE = 0.05-0.20 and $n_0 = 10000$ cm$^{-3}$ with \SFE = 0.09-0.20, respectively. Even when we refer to them as "worst model", they still show a age-mass segregation. We also find similar low value in SEG sample for $n_0 = 8000$ cm$^{-3}$ with \SFE = 0.06-0.20. 

\begin{figure*}
\includegraphics[width=\textwidth]{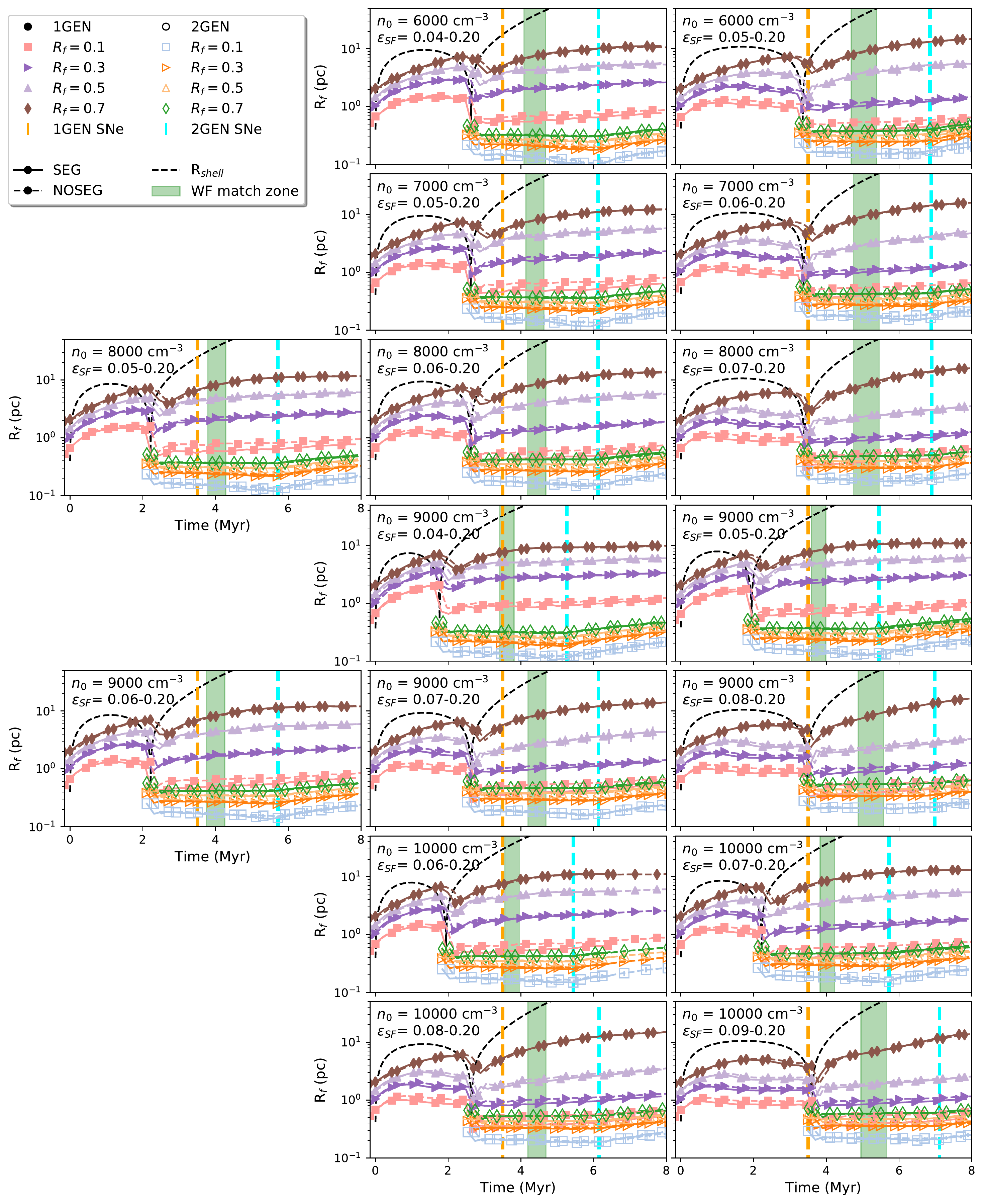}
\caption{Lagrangian radii ($R_f$) vs time. The first generation (1GEN) is denoted by filled symbols and the second generation (2GEN) is denoted by empty symbols.  If the simulations start with mass segregation (SEG) or not (NOSEG) is represented by solid lines or dashed lines, respectively. Shell radius, matching zone and SNe times are as Fig. \ref{fig:rmasm55SEGNOSEG}. Each $R_f$ colour is given in the legend.}
\label{fig:rlagm55SEGNOSEG}
\end{figure*}

\subsection{All stars distribution}
\label{sec:allstarsdist}
We also study the spatial location of the whole stellar distribution with different Lagrangian radii ($R_f$). In specific, we use $R_f =$ 0.1, 0.3, 0.5 and 0.7 separated by generation. We show the results of $R_f$ in Fig. \ref{fig:rlagm55SEGNOSEG}, where 1GEN is represented by filled symbols and 2GEN by empty symbols. Each $R_f$ has a different symbol and colour described in the legend. SEG and NOSEG simulations are represented by a solid and a dashed line, respectively. The shell radius, the matching zone and the SNe beginning are represented as before. In order to do not over-plot the symbols, we shift the SEG and NOSEG results in $\pm 0.10$ Myr resulting in SEG information first. 

As we work with Plummer distribution, SEG and NOSEG mass distributions show initially the same values for the different $R_f$. After this point, we observe that the global evolution followed by both stellar distributions is very similar, being difficult to make a difference without the shift applied to every snapshot. At the moment of the (re)-collapse, the 1GEN stars are not immediately travelling inwards, as it is also observed when only massive stars are analyzed because of the strong gravitational potential produced by the high gas density towards the centre. The different $R_f$, for the older stellar component, only shrink shortly after the new starburst, as the stars need time to change their velocities that were heading outward. After a small contraction, the expansion is resumed until the end of the simulation. The effect of SNe for the old star generation is not appreciable due to the larger expansion produced by the cloud dispersal. For 2GEN, the cloud expansion is not producing big changes in the stellar distribution as they show roughly constant values after the initial contraction due to our initial virial state, until the point when SNe start when in most cases, a new rate of expansion is observed. The behaviour described above is valid for all our sample. About the final star locations, only until $R_f = 0.1$ of 1GEN (pink filled squares) and in some cases $R_f = 0.3$ (purple right triangles) can be comparable with $R_f = 0.7$ 2GEN (green empty diamonds). The rest of 1GEN $R_f$ are always further away from the main concentration of stars. For the case of $R_f= 0.3$ 1GEN can only be comparable with 2GEN $R_f$ when \SFE $\geq 0.08$, being deeper as the 1GEN stellar mass increases. The shell radius (black dashed line) is always larger than the bigger $R_f$. It is only smaller when the (re)-collapse phase is reached, but it reaches a larger position very fast after the second starburst. Every panel shows that the new start cluster which is representing R136 is more concentrated than the old stellar component during the whole simulation so during the green zone when \textsc{warpfield} matches the other observables the $N$-body simulations also match the observed stellar distribution.

\begin{figure*}
\includegraphics[width=0.98\textwidth]{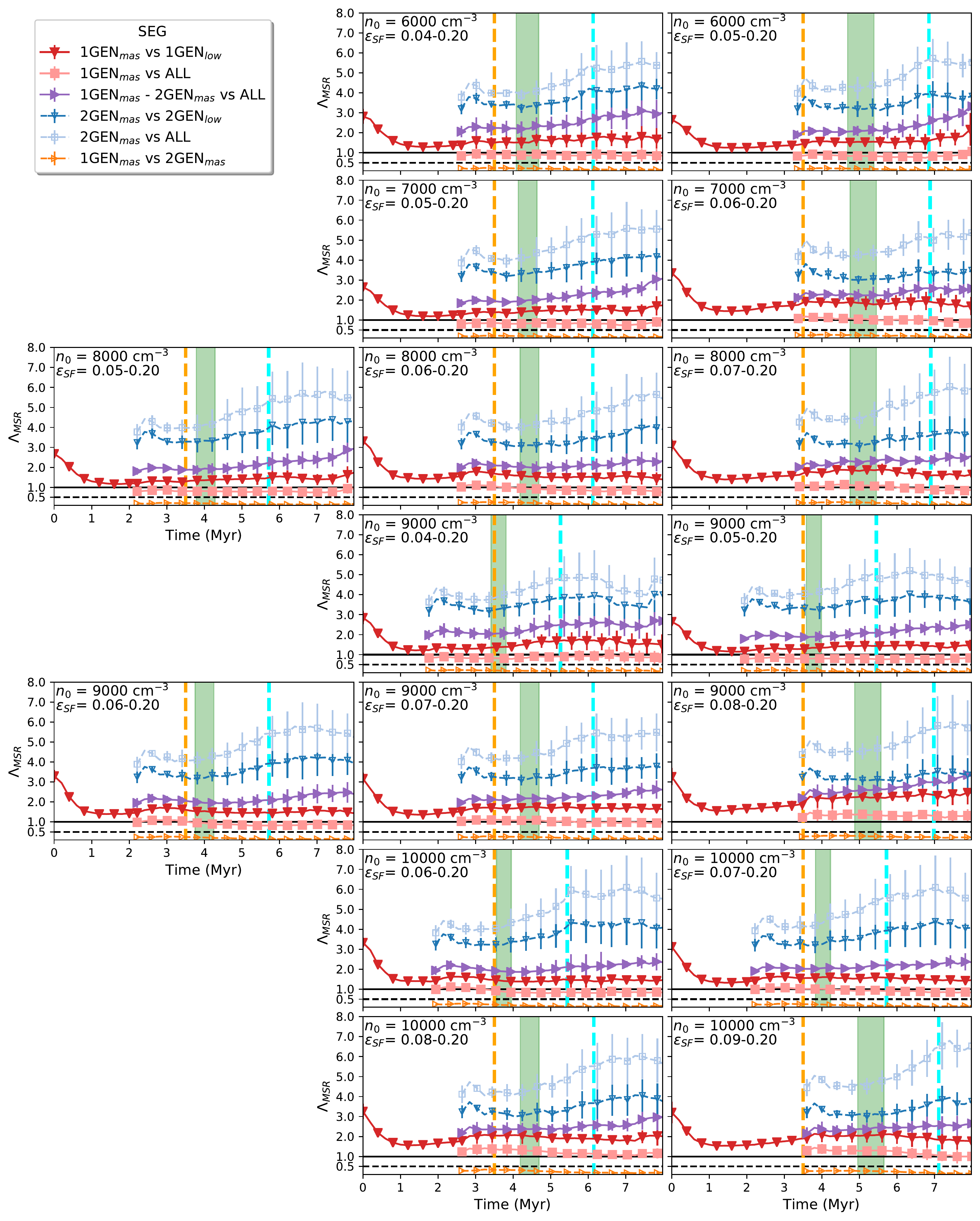}
\caption{Mass segregation (\LMSR) evolution vs time for simulations starting with mass segregation (SEG) and panels ordered as before. Massive stars from the first generation  (1GEN) or second generation (2GEN) are denoted as 1GEN$_{mas}$ and 2GEN$_{mas}$, respectively. Low mass stars are referred as 1GEN$_{low}$ or 2GEN$_{low}$. The rest of the stars excluding the sample of comparison are refereed as ALL. The solid black line shows a value of \LMSR $= 1$ and the dashed black line shows a value of \LMSR $= 0.5$. The vertical orange and blue lines indicate when the first event of SN is taking place for 1GEN and 2GEN respectively. The symbols and colours for each comparison sample are shown in the legend.}
\label{fig:lambdam55SEG}
\end{figure*}

\begin{figure*}
\includegraphics[width=\textwidth]{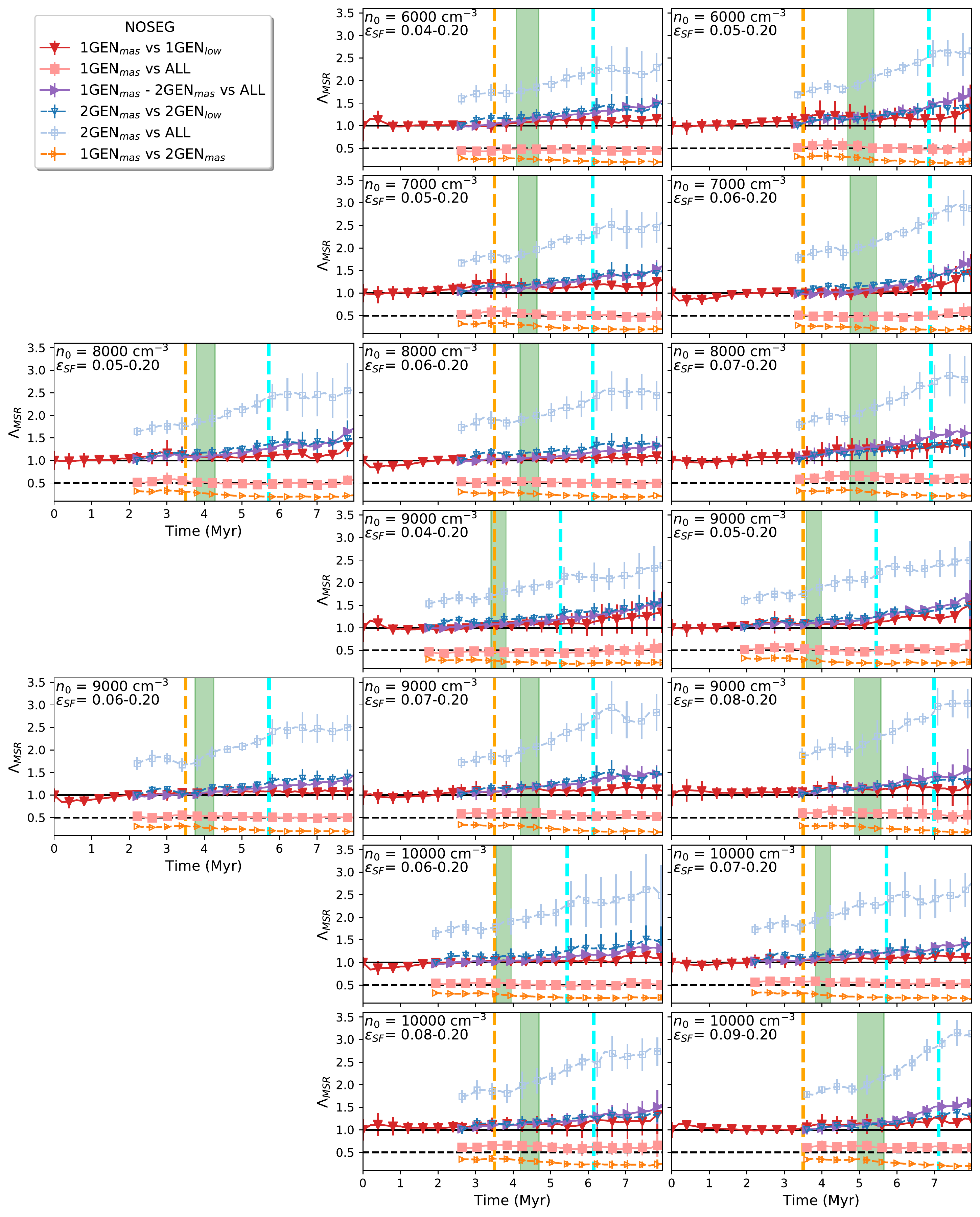}
\caption{Same as in Fig. \ref{fig:lambdam55SEG} but now for simulations starting without mass segregation (NOSEG).}
\label{fig:lambdam55NOSEG}
\end{figure*}

\subsection{Mass segregation}
We use the \LMSR to compare different combinations of star samples from 1GEN, 2GEN or mixed. We consider SEG and NOSEG models separately as we find large differences compared to the analysis above. We measure the mass segregation ratio for the following six combinations:
\begin{itemize}[left= 0pt]
    \item Comparison of first generation massive stars (1GEN$_{mas}$) with the first generation low mass stars (1GEN$_{low}$) (red down filled triangles).
    \item Comparison of 1GEN$_{mas}$ with the rest of the stars (ALL) (pink filled squares).
    \item Comparison of 1GEN$_{mas}$ together with the second generation massive stars  (2GEN$_{mas}$)  with ALL (purple filled right triangles).
    \item Comparison of 2GEN$_{mas}$ with second generation low mass stars (2GEN$_{low}$) (blue down empty triangles).
    \item Comparison of 2GEN$_{mas}$  with ALL (light blue empty squares).
    \item Comparison of 1GEN$_{mas}$ with 2GEN$_{mas}$ (orange empty right triangles).
\end{itemize}

For the SEG sample, we show in Fig. \ref{fig:lambdam55SEG} the previously described combination of \LMSR evolving in time with the panels following the same order as the previous plots. As before the moments when the first SN takes place for 1GEN and 2GEN are indicated by orange and blue vertical dashed lines respectively. The highest level of mass segregation introduced in this sample is detected for the initial conditions at 0 Myr. After this, the gas expulsion occurs and the clusters expand, which produces a reduction in the level of mass segregation. We observe that closely before the second starburst, when the cloud is collapsing and slowly bringing stars from outer locations, the level of mass segregation improves in a small degree and around this value is where it stabilizes and remains until the end of the simulation. The rest of the combinations can only start to be measured after the (re)-collapse as they include stars from 2GEN and our observations are based in comparison to the sample just described. Taking again the 1GEN massive stars but now compared to the rest of the stars (1GEN$_{mas}$ vs ALL) shown as pink filled squares, a lower \LMSR is measured with values close to one or at least always below the comparison sample. All massive stars compared to the rest of the stars results (1GEN$_{mass}$ - 2GEN$_{mass}$ vs ALL) are shown with purple right triangles. In this sample, the starting level of mass segregation is higher (\LMSR $\gtrsim$ 2), which is followed by a decrease but always higher than the comparison sample. The youngest massive stars compared to their respective low mass stars (2GEN$_{mas}$ vs 2GEN$_{low}$) are shown with empty blue down triangles. The initial level of \LMSR starts with the level of mass segregation introduced as an initial parameter, which is reduced as before, but not as much, then it oscillates and finishes at the end of the simulation with a value close to the initial. The massive stars of 2GEN are compared with the rest of the stars (2GEN$_{mass}$ vs ALL) and the results are shown with light blue empty squares. \LMSR shows higher values as it is including stars from 1GEN which are more expanded than the youngest low mass stars, then it oscillates as the previous sample finishing with values higher than the initial. In some of the cases, the final \LMSR can be similar to the initial if 1 $\sigma$ error is taken into account. The last combination is the most important for the goal of this work as it describes if the older massive stars are more expanded than the second generation. We denote this comparison as 1GEN$_{mas}$ vs  2GEN$_{mas}$ and it is shown as orange empty right triangles. We observe a very stable value of inverse mass segregation for all the cases remaining close to the level of mass segregation at the moment when the second stellar generation is introduced to the simulation. At the end of the simulation, we measure values of \LMSR comparable or even less to the initial. The moments of the first SNe for each generation are shown with a dashed vertical orange and blue line respectively. The SNe of the 1GEN are not producing a big change on the value of \LMSR but for the cases which include 2GEN as the comparison sample alone, the SNe are producing instabilities on the \LMSR parameter producing that the rate of increase is reduced and in some cases, when more SNe events are possible, even a decreasing trend towards the end. At the moment when the other observational parameters match the values in the literature (green zone), for all cases, the youngest stellar generation, which is representing R136, is more concentrated than the older stars. 

For the NOSEG sample, we show the result for the same combinations described before in Fig. \ref{fig:lambdam55NOSEG}. Initially, we find \LMSR = 1 as we set up the simulation. Some small decreases or increases are observed but quickly returning to one. After the (re)-collapse, small increases are observed in \LMSR, with the simulations ending at values slightly above one, with maximum values during the simulation typically \LMSR $\sim$ 1.5. 
\LMSR remains close to 1.0 as a result of the fairly uniform global expansion of the cluster (see Sec. \ref{sec:allstarsdist}). From this, we may infer that massive stars are scattered almost at the same level as the low mass stars. When we start to measure 1GEN$_{mas}$ vs ALL after the (re)-collapse, we observe that \LMSR < 1, as the old stellar component has already expanded, it finds a new cluster more concentrated in the centre and in consequence \LMSR detects inverse mass segregation. The level of mass segregation remains stable until the end of the simulation with a value of $\sim$ 0.5. The massive stars of both generation compared with the low mass component shows as 1GEN$_{mass}$ - 2GEN$_{mass}$ vs ALL starts with a \LMSR $\sim$ 1.0 but continuously increasing with values closely above the \LMSR for 1GEN$_{mas}$ vs 1GEN$_{low}$. At the end of the simulation, we detect a minimum mass segregation of 1.5 and taking into account 1 $\sigma$ error a few maximum values of 2. The 2GEN$_{mas}$ vs 2GEN$_{low}$ comparison shows initially a \LMSR = 1.0, as we set up followed by a slow increase until the end of the simulation following closely the previous sample overlapping in the average level of mass segregation finishing with similar values. Comparing all the stars with the massive stars of the new cluster, as 2GEN$_{mass}$ vs ALL, an initial level of mass segregation of at least 1.5 is detected followed by a continuous increase until the end of the simulation. The final \LMSR at 8 Myr are always larger than 2 with most of the cases showing \LMSR $\sim$ 2.5 and a few maximum average values of 3. The comparison of the old and young massive component defined as 1GEN$_{mas}$ vs  2GEN$_{mas}$ shows inverse mass segregation with \LMSR < 0.5 during the whole time of measurement. This value slowly decreases, reaching a level of mass segregation even smaller than the initial. The SNe effect follows the same description as the SEG case showing also instabilities from the moment the SNe for the 2GEN start to take place. At the moment when this last comparison sample is inside the green zone, all the cases show inverse mass segregation as we are aiming to achieve.

\begin{figure}
\includegraphics[width=\columnwidth]{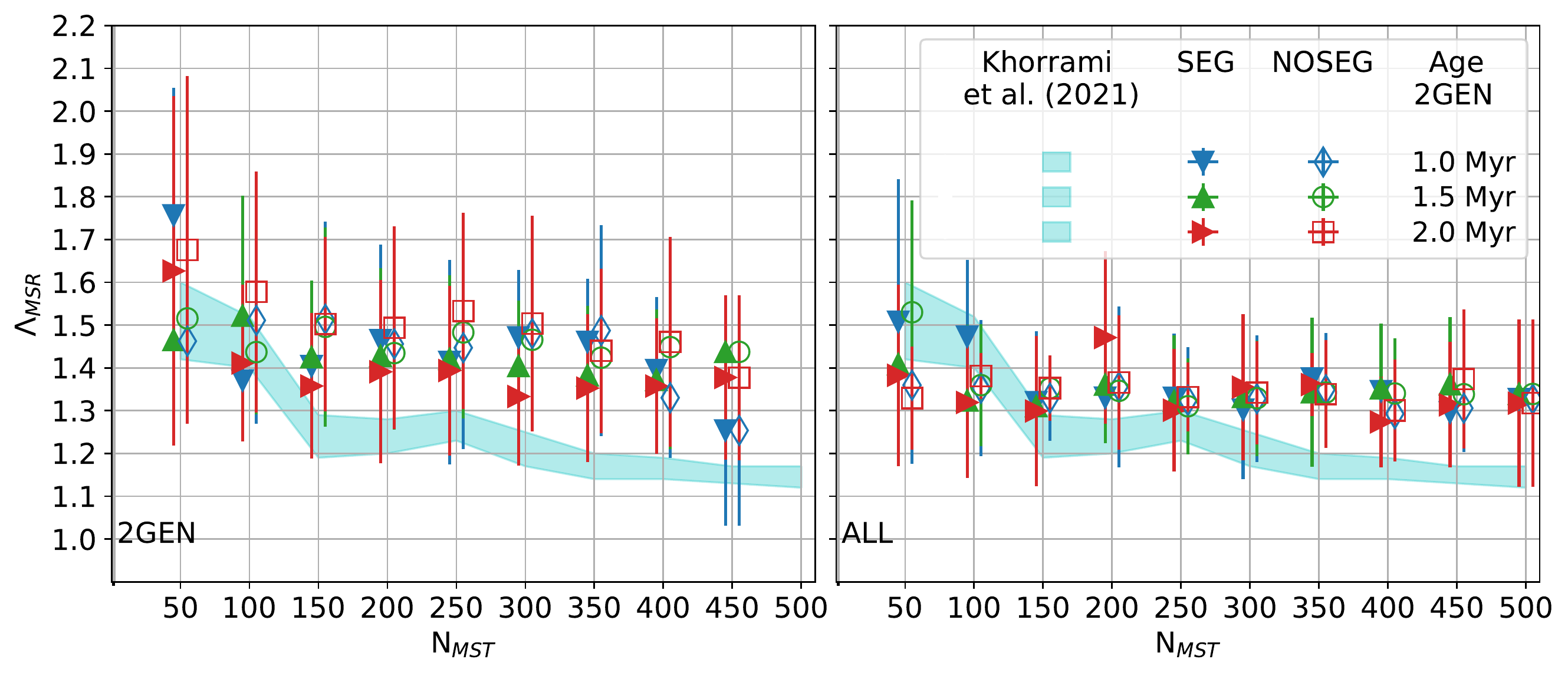}
\includegraphics[width=\columnwidth]{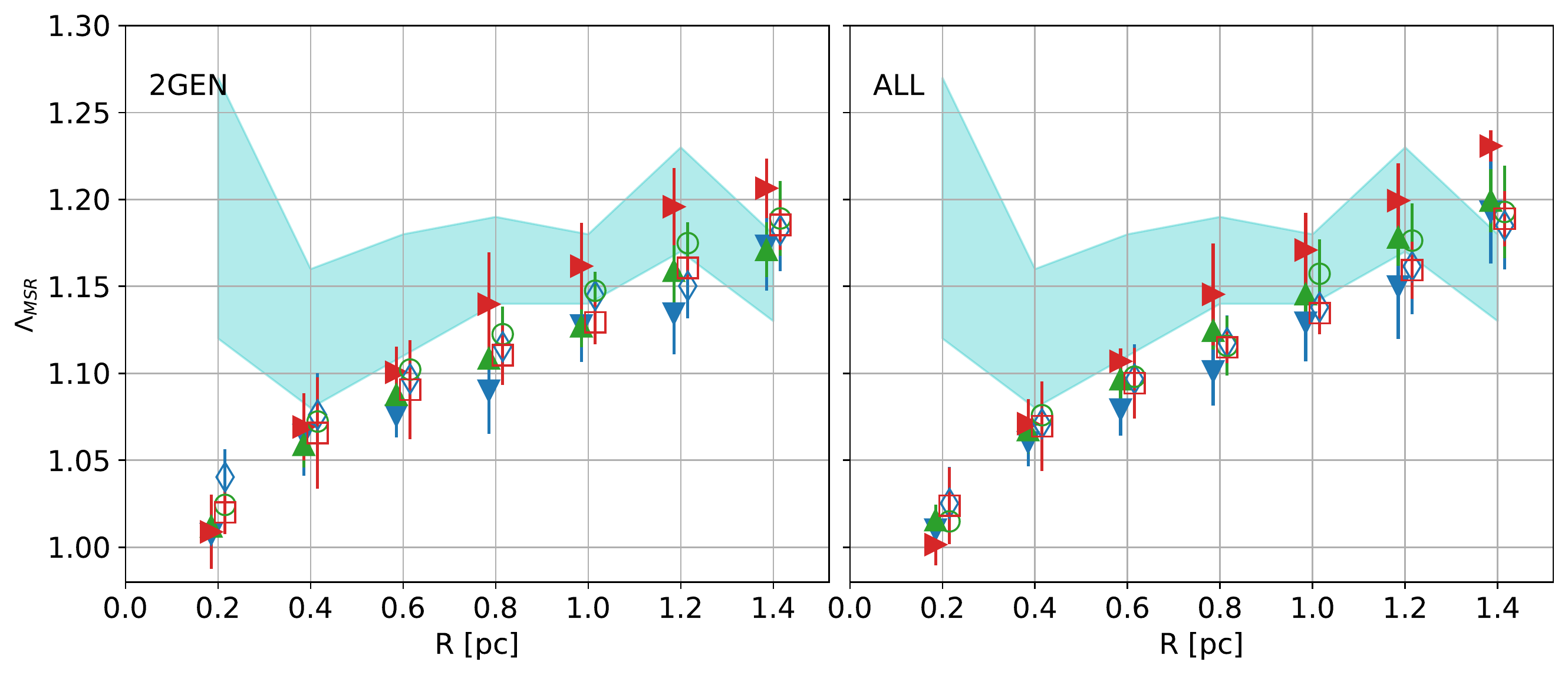}
\caption{Top panel shows the level of mass segregation (\LMSR) measured for different sample sizes of chosen random stars ($N_\text{MST}$). The bottom panel shows \LMSR for different radii. The left and right panels show the results excluding (2GEN) or including (ALL) the old stellar component, respectively. The cyan zones are the observational values from K2021. The different time snapshots and initial mass segregation are indicated by different symbols as the legend denotes. The initial conditions for this case are a cloud of $n_0$ = 10000 cm$^{-3}$ and star clusters according to $\epsilon_{SF}$= 0.07-0.20.}
\label{fig:seg_comp_obs}
\end{figure}

\begin{figure}
\includegraphics[width=\columnwidth]{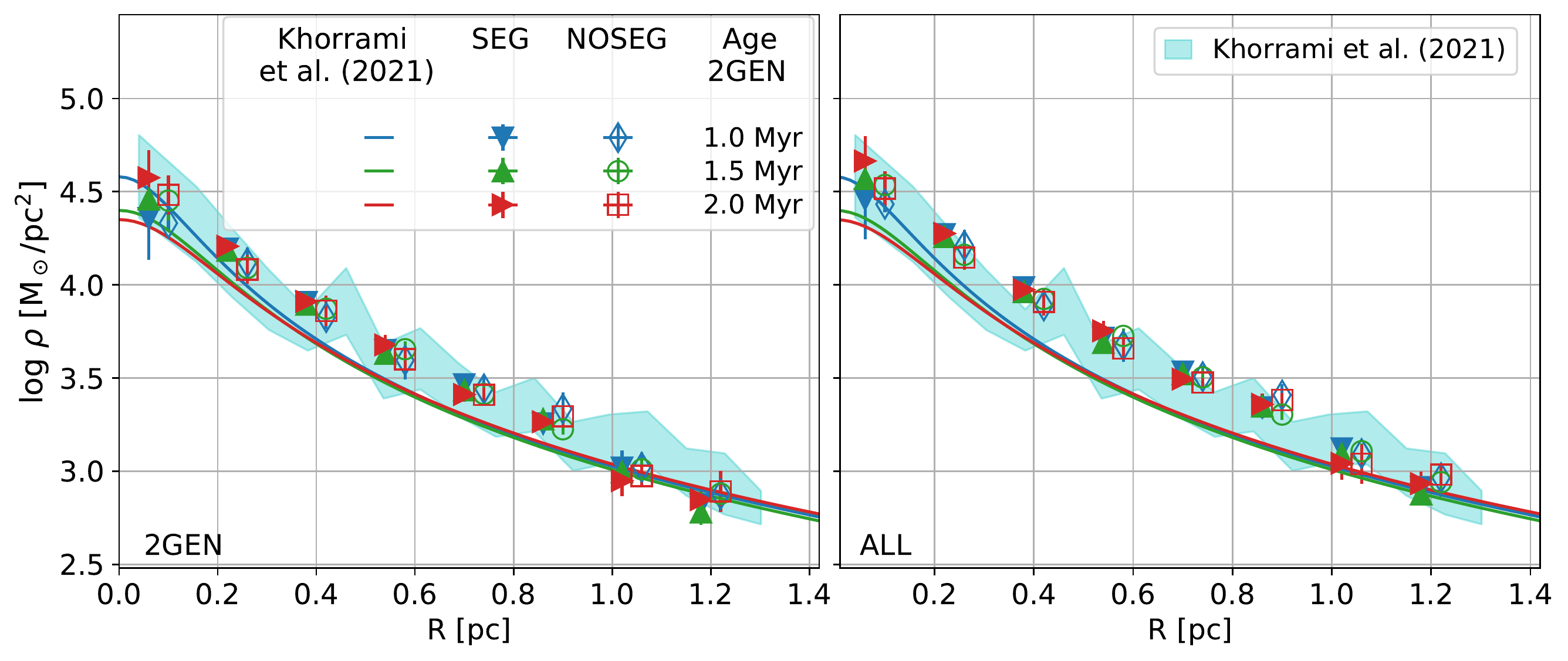}
\includegraphics[width=\columnwidth]{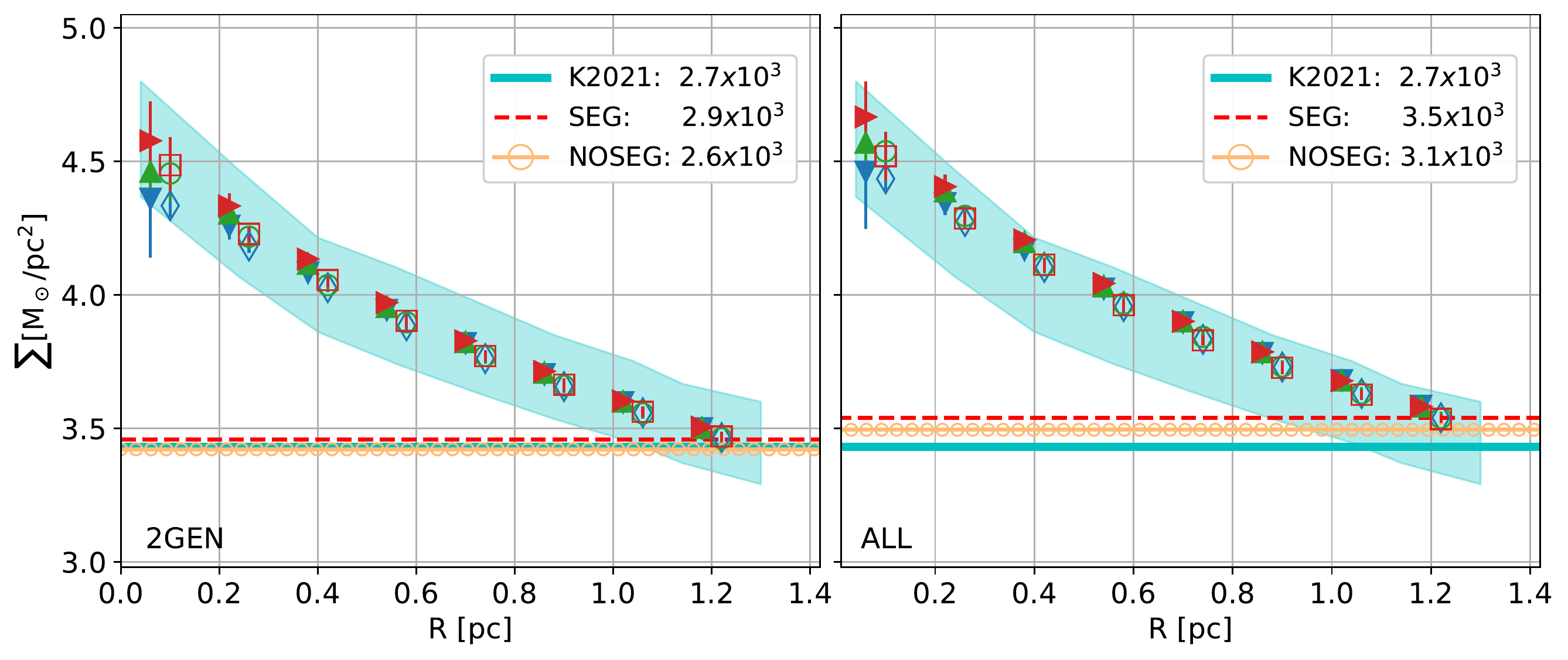}
\includegraphics[width=\columnwidth]{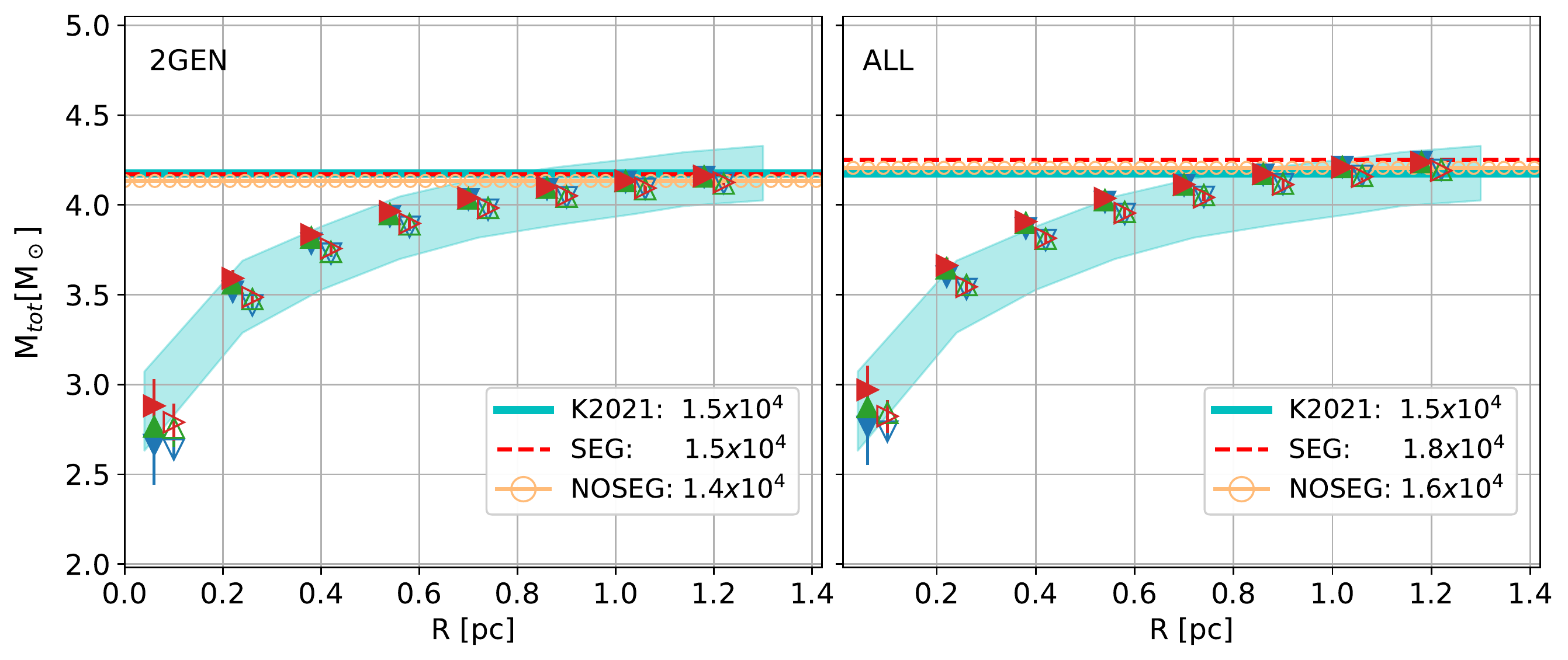}
\caption{Top panels show the projected mass density ($\rho$) of the central zone. The different curves show the fitting lines from the observational study (K2021). The central panels show the surface density ($\sum$) within a given radius. The bottom panels show the total stellar mass ($M_\text{tot}$) within a given radius. The left and right panels show the results excluding (2GEN) or including (ALL) the old stellar component. The cyan lines in the central and bottom panels are the observational values from K2021 and the red line and orange circles are the respective values for simulations starting with mass segregation and not, respectively. The different time snapshots and initial mass segregation are indicated by different symbols as the legend denotes. The initial conditions for this case are $n_0$ = 10000 cm$^{-3}$ and $\epsilon_{SF}$= 0.07-0.20.}
\label{fig:R_rho_surf_Msum}
\end{figure}

\section{Observational measurements}
We compute the \LMSR and density profile of the central zone of our simulations for snapshots at 1.0, 1.5 and 2.0 Myr, trying to match with the observational measurements done by K2021. In our simulations, we have accurate positions, masses and ages for all the particles which for an observational study is not achievable. K2021 presented a detection mass sensitivity which is low in the central zone and it is increasing towards the outer parts. We blind our sample according to the probability of detection presented by them (see Figure 11, bottom panel in K2021). In our simulations, we have two stellar components which would be difficult to distinguish directly in an observation, especially in the small and concentrated central zone. In order to study both cases, we proceed excluding and including the old stars. We find no significant differences between our models, hence, we present the best match which corresponds to the simulations starting with a cloud of $n_0$ = 10000 cm$^{-3}$ and initial clusters with $\epsilon_{SF}$= 0.07 followed, after the (re)-collapse, by our representation of R136 with $\epsilon_{SF}$= 0.20. We present our results using the same plots and units presented by K2021 for a direct comparison. 

\subsection{Central mass segregation}
We measure the \LMSR parameter following the methodology in K2021 including their completeness limitation and observational biases. In Fig. \ref{fig:seg_comp_obs}, top panels, we show the \LMSR calculated for different sample sizes of chosen random stars ($N_\text{MST}$). The cyan zone represents the 1 sigma range from K2021. In the left panel, where the old stars are excluded, our simulations show a flatter trend than the observational results. The central values of \LMSR match the central zone for some cases only for $N_\text{MST} \leq 100$ and for larger $N_\text{MST}$ we can only reach the cyan zone through the 1 sigma error.  In the right panel, where we include the stars from the old component, shows an even flatter curve, with central values below and above the cyan zone. For this case, some of the central \LMSR values are matching for $N_\text{MST} \leq 150$. The 1 sigma ranges as well, for most of the cases, reach the observational zone but with less spread as we increase the  $N_\text{MST}$. The differences between the different 2GEN ages or initial mass segregation are small.
In the bottom panels, we measure the level of mass segregation for different radii. We can only match the cyan zone for $R \geq 0.4$ pc, taking into account the 1 sigma error. At a $R = 0.2$ pc, our results show a \LMSR close to 1 but K2021 shows a larger value of mass segregation, being the only radius where we measure the largest differences between our studies as the cyan zone is never reached either excluding or not the older component. The different time snapshots show similar average values or at least intersect the 1 sigma error.

\subsection{Central density profile}
We measure the 2D mass density profiles for a given radius as it was done by K2021 and we summarize the results in Fig. \ref{fig:R_rho_surf_Msum} top panel. The different curves are the mass density profiles shown in K2021 at different estimated ages. Our results are matching the curve close to the centre ($R < 0.2$~pc) and staying slightly above until $R \sim 1$~pc where again match the solid lines. The same behaviour is shown when the old component is excluded or not. We do not observe big differences for any time snapshots or initial mass segregation. The results shown in K2021 are also not matching the curves perfectly as we can see from the cyan zone which denotes the spread of the results measured by the authors. 
In the central panels, we show the surface density for different given radii. We find that our results follow similar curves but the final values show differences. The final mass density found by K2021 is $\sum = 2.7 \times 10^3$ M$_\odot/\text{pc}^2$ and our best match in this case is given in the left panel with small differences as $-0.1 \times 10^3$ M$_\odot/\text{pc}^2$ and $+0.2  \times 10^3$ M$_\odot/\text{pc}^2$ for NOSEG and SEG simulations respectively. In the right panel, where all stars are included, higher surface densities in the order of $\geq +0.4$ M$_\odot/\text{pc}^2$. The results are in both cases inside the cyan zone but when the old component is included the values approach the top limit.
In the bottom panels, we show the stellar mass for given radii. As before, the closest values are observed in the left panels. K2021 estimated a total mass of M$_\text{tot}=1.5 \times 10^4$ M$_\odot$ and our results can match this value for SEG simulations and with a difference of less than 10\% ($-0.1  \times 10^4$ M$_\odot$) for NOSEG simulations. In the right panel, NOSEG simulations also find a close value with a difference of less than 10\% ($+0.1  \times 10^4$ M$_\odot$). Simulations starting with mass segregation enclose more mass, resulting in a value above the observation measurement of $+0.3 \times 10^4$ M$_\odot$. We also observe that in the right panel our results are closer to the top limits of the cyan zone.

\section{Discussion}
In this work we demonstrate that the stellar distribution observed in NGC~2070 is consistent with an older stellar cluster, dynamically relaxed, hosting in its centre a youngest more massive star cluster known as R136. We achieve this through $N$-body simulations coupled with a semi-analytic 1D model for evolution of cloud/cluster systems.

We evolve a molecular cloud initially with $M_\text{cloud}~=~3.16~\times~10^{5}$~M$_\odot$ trying initial uniform densities ($n_0$) between 6000-10000~cm$^{-3}$ holding different star clusters leading in \SFE between 0.04-0.09. We scale the velocities in order to obtain dynamical equilibrium ($\alpha=0.5$). All the combinations shown in this work include clusters which produce insufficient feedback to dissolve the cloud, despite being massive and young. As a consequence, the cloud (re)-collapses and a second starburst occurs. We fix this second star cluster to have a \SFE = 0.20. The last imposition is made in order to have more time to match the ages of both stellar generations and the shell radius. After several attempts exploring the best parameter space to reproduce the observables of R136, we find that a second star cluster starting in dynamical equilibrium is not able to match observations. We explore different $\alpha$, and we find that the best dynamical state for the new stellar component is $\alpha=0.3$ i.e., the second cloud expansion is holding a new cluster that is initially contracting. The dynamics of the second cluster is affected by the dynamics of an expanded older less massive stellar component and the expanding cloud which is removing gravitational potential. We also explore if our results can vary if both stars clusters start with mass segregation or not. In this paper, we only include a summary of the results for the successful $\alpha$. 

We study NGC~2070 as a whole measuring the average distances to the centre for only the massive stars ($M > 20$ M$_\odot$) weighted by their luminosity. We find for all the cases that the new massive stars stay captive closer to the centre and the remaining older massive stars are at further locations as is observed in NGC~2070. It is important to mention that this is not saying that there are no old massive stars close to the centre, in fact, we detect them but they are not that many to reduce the average central distance. We observe that the different stellar components can be easier recognized if the star clusters start with mass segregation as the new massive stars are found much more concentrated than their pairs. At later stages, the expansion of the new stellar component is larger and we achieve this as a consequence of the SNe which have more time to be produced.  
The massive stars which belong to the clusters starting with mass segregation are found, on average, closer to the centre due to their initial imposed configurations and this difference is more visible for the newer massive stars. 

We continue studying the Lagrangian radii of both stellar components. We observe that independent of the initial level of mass segregation, the old stellar component is always found more extended than the new cluster. The initial contraction for the second stellar generation because of our imposed virial ratio is visible along all the layers. Its expansion is stronger when the new SNe start to be produced at later stages of our simulations but does not influence the matching scenarios as this occurs later than the moment when all the observable are intersecting. The clusters which start with mass segregation are slightly more concentrated than their pairs and this is observable through our complete sample. 

To quantify our model in a physical way, we measure the level of mass segregation for the different cases using the \LMSR parameter. As the new stellar component is more concentrated than the old star cluster, we expect to find a \LMSR $< 1$, when we compare the massive stars from the older generation with the new massive stars. We show the results in separated plots this time as the initial \LMSR differ highly when we start with mass segregation (\LMSR $>> 1$) or not (\LMSR $\sim 1$). The old cluster which starts with mass segregation loses this configuration due to the cloud expansion and at the moment of the inclusion of the new star cluster, their \LMSR $ \gtrsim 1$ i.e., a cluster without mass segregation as their pairs which at the same moment also exhibit the same distribution. After the inclusion of the new clusters, the comparison between the different samples shows similar behaviours. At the moment when the observables in \textsc{warpfield} are matched every combination of initial conditions show a value \LMSR $\sim 0.5$ for the cases without initial mass segregation or even less when we start with segregated clusters. Along our whole parameter space, the NGC~2070 stellar configuration is detected regardless of any of the conditions on the initial conditions here used. 

We also compare to the study of K2021, who present observations of the central region of NGC~2070 where the younger cluster R136 is located and discuss the resulting radial mass segregation and density profiles. We proceed as closely as possible to their approach and we find a good match with the observations. Unlike the results on mass segregation previously exposed, we exclude any star with a central distance larger than 1.4 pc, as the observational study has done. We can closely match the descending trend for the \LMSR parameter as we increase the size of the sample (N$_\text{MST}$). Using only the new stellar component which in our work represent R136, we do not match the exact central values for every case, but our 1 sigma error bars are always close to the observational values. Adding the remaining stars from the old stellar component in this zone, we can match the central values. This means some older stars can be also contaminating the observational study. We measure \LMSR at different central distances and we find an increasing trend. The increasing trend is also found by K2021, with only one exception at a central distance of 0.2 pc. This discrepancy is found with our without the old stellar component. Our central values are slightly below the observational results but always intersected by 1 sigma error bars. It has been detected, very close to the R136 centre, a star with a mass $\sim$ 300 M$_\odot$, but in this work, we did not extend our initial mass function further than  $\sim$ 120 M$_\odot$ as expecting to find this massive star in our simulation also very close to the centre can be challenged due to the stochastic dynamical interactions. This star in the very centre taken into account by observers improves the value of mass segregation at R $\leq 0.2$ pc showing the biggest difference between our works. In the referenced study, the values of mass segregation cover a range of 1.0 $\leq$ \LMSR $\leq 1.28$ which is very small for this parameter and it can vary easily depending on the random star selection \citep{2009MNRAS.395.1449A}. For radial density profiles, we find good agreement between observations and our simulations. We can match very accurately the observational values with our central values. We can only match these radial profiles if at the moment of the introduction of the new star cluster, instead of being in equilibrium, it is contracting. We try with different virial ratios ($\alpha \leq 0.5$) and the best agreement with observations is $\alpha=0.3$ which are the results presented in this work. We find this independent of the initial mass segregation, initial cloud density, and star formation efficiency pairs.

\section{Conclusions}
We conclude that an evolving molecular cloud with an initial mass of 3.16$\times 10^5$ M$_\odot$ giving birth to two stellar generations can well reproduce the observational characteristics of the central region of 30 Doradus in the Large Magellanic Cloud. Our model of an older first-generation star cluster with a mass between 1.26$\times 10^4$ M$_\odot$  and 2.85$\times 10^4$ M$_\odot$, starting in virial equilibrium, followed by a younger second-generation cluster of $\approx 6.32\times 10^4$ M$_\odot$, starting contracting with a virial ratio of 0.3, can match the stellar configuration observed in NGC~2070 consisting in an old expanded cluster hosting in its centre a youngest more massive star cluster known as R136. The resulting new stellar component shows close agreement with mass segregation observations of R136 excepting the very central zone ($R < 0.2$ pc) where a $\sim$ 300 M$_\odot$ is located which has been not included in this work. Whether we include remnants from the old component or not, our simulations match the density profile of the central zone of NGC~2070. Therefore, this result is independent of the probable contamination by old stars in K2021.

We caution that there may be other configurations that lead an equally good match to the observational constraints. 
The approach presented here is kept simple in order to allow for the investigation of a large parameter space. Subsequent studies based on complex and computationally more expensive 3D radiation-hydrodynamic simulation can use our best fit model as starting point.

We observe that the second stellar generation, representing R136, remains more concentrated than the first generation, which can be well understood as a natural outcome of the stellar dynamical evolution in the time-varying potential. We mention, that the \textsc{warpfield} model could in principle produce more massive star clusters that also match the ages and shell radius of NGC~2070, however, in these cases the spatial distribution of stars is typically too extended to be compatible with the observational constraints.

\section*{Acknowledgments:}  We acknowledgement support from ANID (CONICYT-PFCHA/Doctorado acuerdo bilateral DAAD/62170008) and from  from the German Academic Exchange Service (DAAD) in funding program  57395809. The authors acknowledge support by the state of Baden-W\"urttemberg through bwHPC and the German Research Foundation (DFG) through grant INST 35/1134-1 FUGG providing the computing power to conduct the simulations suite presented here. We also thank the DFG for financial aid via the Collaborative Research Center SFB881 (Project-ID 138713538) {\em The Milky Way System} in subprojects A1, B1, B2, and B8, and we acknowledge encouragement from the Heidelberg Cluster of Excellence EXC 2181 (Project-ID 390900948) {\em STRUCTURES: A unifying approach to emergent phenomena in the physical world, mathematics, and complex data} funded by the German Excellence Strategy.

\section*{DATA AVAILABILITY STATEMENT}
The data of the full set of simulations (see Tab. \ref{tab:tabsims1}) presented in this article will be shared on reasonable request to the corresponding author.

\bibliographystyle{mnras}
\bibliography{bibtex} 

\label{lastpage}

\end{document}